\documentclass[11pt, a4paper, twoside]{report}

\usepackage[english]{babel}
\usepackage[utf8x]{inputenc}
\usepackage[T1]{fontenc}

\usepackage[a4paper,top=2.5cm,bottom=2.5cm,left=2.5cm,right=2.5cm,marginparwidth=1.75cm]{geometry}

\usepackage{amsmath}
\usepackage{graphicx}
\usepackage[colorinlistoftodos]{todonotes}
\usepackage[colorlinks=true, allcolors=blue]{hyperref}
\usepackage{float}
\usepackage{setspace}

\title{Matching Models for\\ Graph Retrieval}
\author{Yash Jain  Chitrank Gupta}

\RequirePackage{tikz}
\RequirePackage{pgfplots}
\pgfplotsset{compat=1.15}
\RequirePackage{pgfplotstable}
\usetikzlibrary{shapes,arrows,positioning}
\RequirePackage{fp}

\RequirePackage{amsmath,amssymb}
\RequirePackage{xcolor}
\RequirePackage{graphicx,soul,colortbl,subcaption}

\RequirePackage[normalem]{ulem} 
\colorlet{redgray}{red!35!gray}
\colorlet{greengray}{green!35!black}
\colorlet{bluegray}{blue!50!gray}

\usepackage{latexsym}
\usepackage{mathrsfs}
\usepackage{bbm}
\usepackage{etoolbox}
\usepackage{algpseudocode}
\usepackage{amsmath,cleveref,amssymb}
\usepackage[normalem]{ulem}
\usepackage[ruled,vlined]{algorithm2e}
\RequirePackage{Definitions}

\begin{document}

\begin{titlepage}

\newcommand{\HRule}{\rule{\linewidth}{0.5mm}} 


\hspace{4cm}\includegraphics[width=5cm]{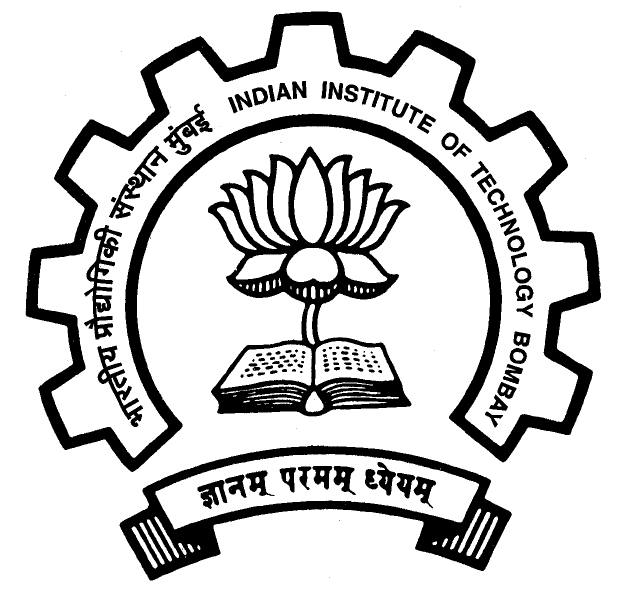}\\[1cm] 
 

\center 


\textsc{\LARGE Research Report Fall 2020}\\[1.5cm] 
\textsc{\Large Indian Institute of Technology Bombay}\\[0.5cm] 
\textsc{\large Department of Computer Science and Engineering}\\[0.5cm] 

\makeatletter
\HRule \\[0.4cm]
{ \huge \bfseries \@title}\\[0.4cm] 
\HRule \\[1.5cm]
 

\begin{minipage}{0.4\textwidth}
\begin{flushleft} \large
\emph{Author:}\\
  Yash Jain\\
  \texttt{yashjain@cse.iitb.ac.in}
  Chitrank Gupta\\
  \texttt{baekgupta@cse.iitb.ac.in}
\end{flushleft}
\end{minipage}
~
\begin{minipage}{0.4\textwidth}
\begin{flushright} \large
\emph{Supervisors:} \\
Prof. Soumen Chakrabarti \\ 
\texttt{soumen@cse.iitb.ac.in} \\
Prof. Abir De \\
\texttt{abir@cse.iitb.ac.in}
\end{flushright}
\end{minipage}\\[2cm]
\makeatother



{\large \today}\\[2cm] 

\vfill 

\end{titlepage}

\begin{abstract}
Graph Retrieval has witnessed continued interest and progress in the past few years. In this report, we focus on neural network based approaches for Graph matching and retrieving similar graphs from a corpus of graphs. We explore methods which can soft predict the similarity between two graphs. Later, we gauge the power of a particular baseline (Shortest Path Kernel) and try to model it in our product graph random walks setting while making it more generalised.

\end{abstract}


\tableofcontents
\listoffigures
\begingroup
\let\clearpage\relax
\listoftables
\endgroup

\chapter{Introduction}

\section{Objectives}
The objective of this study was to:
\begin{itemize}
    \item Improve the performance of the SIGIR paper "Deep Neural Matching Models for Graph Retrieval"  \cite{10.1145/3397271.3401216} by incorporating trainable random walks in GxNet

    \item Implement new graph kernel baselines apart from the GxNet paper \cite{10.1145/3397271.3401216}
    \item Survey new graph datasets for robust testing
    \item Develop novel shortest path product graph approach to generalise Shortest path kernel \cite{1565664} in random walk setting
\end{itemize}


\section{Contributions}
During this study, we developed a thorough understanding of graph based information retrieval systems. We learnt the pitfalls in various graphs based IR systems and tried to improve the performance of GxNet by incorporating trainable random walks in its architecture. 
Further, we tried to generalise the Shortest Path Kernel approach by integrating its ideas into product graph regime to match its  performance with Shortest Path Kernel in random walk framework on several datasets. 


\chapter{Matching Models for Graph Retrieval via Deep Learning}
\section{Introduction}
The graph retrieval problem is to return ‘relevant’ or ‘good’ response graphs from a corpus, given a query graph. The solution to this problem requires scoring (and thereby rank) the corpus graphs with respect to a query graph using some technique. Though in each dataset, the concept of relevance is different, i.e. either semantic or structural and hence, a single heuristic will not work for both types of such datasets. Most prevalent graph kernels quantify the structural similarity between graphs, providing high similarity scores to isomorphic graphs. That's why this check of similarity between graphs is also called isomorphism check. On the other hand, in many other applications like drug active-sites identification or question answering from documents, we may want a corpus graph to be prompted relevant if a subgraph of the corpus graph is isomorphic to the query graph. Thus, GxNet proposes an E2E supervised approach which implicitly learns such isomorphic subgraphs.

\section{Notation}
Given a set of query graphs $\Qcal=\left\{G_q=(V_{q},E_{q}, F_q, T_q)\right\}$  and a set of corpus graphs   $\Ccal=\{G_c=(V_{c},E_{c}, F_c, T_c)\}$ .  Each query graph $G_q$ is associated with a set of \emph{good} (relevant) corpus graphs 
\begin{align}
 \Ccal_{q+} &=\{G_c\in\Ccal|\text{Relevance}(\Gcal_q,\Gcal_c)=1\}
\intertext{and a set of \emph{bad} (irrelevant) corpus graphs}
\Ccal_{q-} &=\{G_c\in\Ccal|\text{Relevance}(\Gcal_q,\Gcal_c)=0\}.
\end{align}

For each query or corpus graph $G_{\bullet}$ we also observe a set of node features $F_{\bullet}=\{f_u \}_{u\in V_\bullet}$ and a set of edge types $T_{\bullet}=\{t_e\}_{e\in E_\bullet}$, where $\bullet$ may be one of $q$ and~$c$. The meaning of the underlying node features and edge types vary across datasets.

\paragraph{Product graph}
For each pair of query $G_q$ and corpus graph $G_c$, we construct the tensor product graph ${\Gcal}^{qc}=G_q \times G_c$
:
\begin{align}
{\Gcal}^{qc}=G_q \star  G_c  =({\Vcal}^{qc}, {\Ecal} ^{qc}, {\Fcal} ^{qc}, {\Tcal} ^{qc}), \; \text{where} \\
\notag {\Vcal}^{qc}\!=\! V_q\!\times\! V_c, \; {\Ecal}^{qc}\!=\! E_q\! \times\! E_c, \; 
{\Fcal}^{qc}=F_q \times F_c, \; {\Tcal}^{qc}&= T_q \times T_c.
\end{align}
Edges are included in the product graph $\Gcal^{qc}$ according to the rule
\begin{align}
&((u,u'),(v,v')) \in {\Ecal}^{qc} \!\iff\! (u,v) \in E_q \ \text{and} \ (u',v') \in {E_{c}};
\end{align}
the node feature for a node $(u,u')$ in $\Gcal^{qc}$ are set as follows:
\begin{align}
&f_{(u,u')} = (f_u,f_{u'}) \ \forall u\in G_q, u'\in G_c;    
\end{align}
and the edge labels in $\Gcal^{qc}$ are set as follows:
\begin{align}
& \hspace*{-2mm}t_{(u,u'), (v,v')}= (t_{u,v},t_{u',v'})\ \text{if } (u, v)\in  E_q \ \text{and} \ (u', v')\in  E_c
\end{align}

\section{Method}
GxNet computes an embedding $\zb_{u,u'}$ for each node $(u,u')$ in the product graph $\Gcal^{qc}$. In addition, GxNet also computes an embedding $\bm{r}_{ee'}$ for each edge in the product graph based on the edge features $\tb_e, \tb_{e'}$. More specifically, we have:
\begin{align}
\bm{z}_{u,u'}&=\gb_{\text{node}}(\fb_{u},\fb_{u'})\\
\bm{r}_{e,e'}&=\gb_{\text{edge}}(\tb_e,\tb_{e'})
\end{align}
In the next step, GxNet constructs the embedding vector for each (source node, edge, target node) triple in the product graph $\Gcal^{qc}$,
\begin{align}
\text{denoted} \quad \tau &= \big((u,u')\overset{ee'}{\longrightarrow}(v,v')\big), \\
\text{as} \quad
\hb_{\tau} &= \bm{\rho}(\zb_{u,u'},  \rb_{e,e'},  \zb_{v,v'}),
\end{align}

which in turn gives us a score or scalar  weight of $\tau$, given by 

\begin{align}
s_\tau= \sigma(\Wb_{s,h}^\top \hb_{\tau}),
\end{align}
where $s_{\tau}$ indicates an affinity score     for the edge $\tau$.
Here $\gb_{\text{node}}$, $\gb_{\text{edge}}$, $\bm{\rho}$, $\sigma$ are suitable standard networks with appropriate non-linearities.  The parameter $\theta$ stands for the set of all the trainable parameters in these networks. 
The final score $y_{\theta}(G_q,G_c)$ is calculated by aggregating the scores $s_\tau$ of a Subgraph $S$ in the product graph $\Gcal_{qc}$. The idea is that if $G_q$ and $G_c$ are positively related then there will be a subgraph in both of them (and subsequently $\Gcal_{qc}$) which highly ‘matches’ with each other. The neural networks hopes to learn this latent substructure. 

Since finding such a $S$ is tedious and almost impractical GxNet employs Random walks of length $K$ (a hyperparameter) to sample a part of the product graph and get an aggregated score with the hope of sampling the similar subgraph in $G_q$ and $G_c$.
Formally, for each $k$ $\epsilon$  [$K$], it performs $n$ random walks with restarts in the product graph. The start probability of all nodes is kept equal. At any node, transition probabilities for all edges are equal. After sampling a random walk, it samples the subgraph as the induced subgraph of a random walk. After sampling subgraphs $S_i, i \leq n$ computes the approximate maximum of the aggregated affinity scores in the following way.
\begin{align}
\text{score}_k(G_q,G_c) = 
\sum_{i\in[n]} \dfrac{  \sum_{\tau\in \Scal_i||\Scal_i|=k} s_{\tau} \exp(\sum_{\tau\in \Scal_i||\Scal_i|=k}s_{\tau})}{\sum_{i\le n}\exp(\sum_{\tau\in \Scal_i||\Scal_i|=k}s_{\tau})}
\end{align}

Finally, GxNet compute the average affinity over the different lengths of the walk, penalized by the length of path $k$. Hence, the final confidence score of the relevance of $G_q$ with $G_c$ is
\begin{align}
    y_{\theta} (G_q,G_c)=\frac{1}{K}\sum_{k\in [K]}\frac{\text{score}_k(G_q,G_c)}{k},\label{eq:y}
\end{align}

To train the parameter $\theta$, GxNet uses pairwise ranking loss function
\begin{align}
    \underset{\theta}{\text{argmin}} 
       \sum_{q\in\Qcal}\sum_{\substack{G_c\in \Ccal_{q-},\\ G_{c'}\in \Ccal_{q+}}} \!\!\!
       \text{ReLU}\Bigl(\Delta + y_{\theta} (G_q,G_{c}) - y_{\theta} (G_q,G_{c'}) \Bigr)
\end{align}
\hspace{0.6cm} where $\Delta$ is a tuned margin and ReLU is $\max\{\cdot, 0\}$.

\section{Results}
GxNet was evaluated on two datasets, i.e. VQA and SQuAD, and was compared against the baselines GMN and RRWM. Please refer to the table \ref{tab:ch2} for detailed results.

\begin{table}[]
\centering
{%
\begin{tabular}{|l|l|l|l|l}
\cline{1-4}
\textbf{Dataset}                                                      & \textbf{GxNet} & \textbf{GMN} & \textbf{RRW} & \textbf{} \\ \cline{1-4}
\textbf{\begin{tabular}[c]{@{}l@{}}VQA (Clevr)\end{tabular}} & \textbf{0.98}  & 0.38         & 0.65         &           \\ \cline{1-4}
\textbf{SQuAD}                                                        & \textbf{0.34}  & 0.27         & 0.37         &           \\ \cline{1-4}
\textbf{COX2}                                                         & \textbf{0.31}  & 0.35         & 0.37         &           \\ \cline{1-4}
\end{tabular}%
}
\caption{Mean Average Precision Values across various baselines. }
\label{tab:ch2}
\end{table}

Preliminary experiments suggested that GxNet showed great promise in graph neural matching and was highly resilient to noise in the data compared to its counterparts. We will delve in further about GxNet in future chapters.

\chapter{Datasets}
\section{SQuAD \cite{rajpurkar2018know}}
Stanford Question Answering Dataset (SQuAD) is a reading comprehension dataset, consisting of questions posed by crowdworkers on a set of Wikipedia articles, where the answer to every question is a segment of text, or span, from the corresponding reading passage, or the question might be unanswerable. To make it amiable for the graph retrieval task, each natural language sentence was converted into its dependency parse using spacy dependency parser. The corpus graph was made up of the rest of the spans from the reading passage.
\section{COX2}
 COX2 is a molecular dataset where every graph is classified into one of two classes.
\section{PTC}
 PTC is a molecular dataset where molecules are classified according to their carcinogenicity on rodents. This dataset is classified into 4 datasets\- PTC\_FM,PTC\_FR,PTC\_MM and PTC\_MR where M/F stands for male/female and M/R stands for mice/rodent.
\section{MUTAG}
 This dataset is a molecular dataset where molecules are classified according to their mutagenicity on bacteria.
\section{Clevr VQA}
 The Clevr VQA benchmark consists of an image corpus where each image is associated with a scene graph. The nodes of a scene graph have attributes concerning shape, color, texture and size. Edges of a scene graph represent spatial relations like ‘left of’ and ‘behind’. Clevr queries are originally in natural language and usually parsed into a graph. The query graphs were generated by placing five objects randomly on the scene. Object attributes (shape, color, size, material) are sampled uniformly randomly from predefined choices. Next, spatial relations (left, right, front, behind) are computed between object pairs. The label of an edge is a 1-hot vector of length 4, corresponding to the above relations. Positive corpus graphs are generated by adding more objects to the scene whose attributes are decided uniformly randomly, and edges are added by computing relationships between object pairs as before. Negative corpus graphs are sampled from the positive samples of other queries since the probability of a corpus graph matching two query graphs well is very low. To test the resilience of various algorithms, noise is added to the color (RGB ∈ R 3 ) and size (∈ R) node attributes. Each component is in [0, 1], to which an independent Gaussian noise with varying standard deviation was added.
\section{Pascal VOC Keypoints}
 Pascal Visual Object Classification challenge consist of 20 image classes, where each image was parsed into an image graph using keypoints as nodes. The nodes signify the VGG16 features of the bounding box of the part of image local to the keypoint. The dataset was later rejected for the experiments as the node features within the classes does not make sense semantically, i.e., nodes of an image-graph of aeroplane cannot be compared to the nodes of a car but only be compared to an another instance of aeroplance image-graph. This was unsuitable for our proposed task of deep graph matching with a query and a set of corpus graphs.

\chapter{Baselines}
\section{Non-trainable Heuristics based}
\subsection{Random Prediction Baseline}
Here we arbitrarily predict the similarity between the graphs.
\subsection{Graph Kernels}
A kernel is a function that takes two vectors and returns the similarity between these two vectors as the dot product between some high dimensional mapping of the two vectors. The two conditions equivalent to a kernel function are symmetricity and positive semi-definite property. In context of graph kernels, the higher dimensional feature space most often corresponds to the decomposition of the graph into some specific substructures like walks, paths, sub-graphs which may or may not be further refined. Here we consider two different graph kernels- reweighted random walk kernel and shortest path kernel.
\subsection{Shortest Path Kernel \cite{spk}}

This kernel essentially calculates the similarity between the graphs by comparing the two graphs on the basis of the no of shortest paths of some specific length and connecting two nodes of some specific node labels. Formally let $\phi_(n,l_u,l_v)(G)$ denote the number of shortest paths in graph G of length $n$ that exist between two nodes with node labels $l_u$ and $l_v$. Then we have,
\begin{equation}
\begin{split}
    \phi_{(n,l_u,l_v)}(G) =\lvert (u,v): \mathtt{label}(u) &=l_u, \mathtt{label}(v)=l_v \\ & \mathtt{shortest\_path}(u,v)=n \;\forall u,v \in V_G \rvert
\end{split}
\end{equation}
We do this for all such possible $k$,$l_u$ and $l_v$ which can be done in polynomial time thanks to algorithms like Floyd-warshall. Once done this, then to compute the similarity of some graph in the corpus $G_c$ with that a query graph $G_q$ one can do the following. 
\begin{equation}
    \begin{split}
        k_{shortest\_path}(G_q,G_c)= \sum_{(n,l_u,l_v)} \phi_{(n,l_u,l_v)}(G_q) .\phi_{(n,l_u,l_v)}(G_c)
    \end{split}
\end{equation}
One can rewrite this formula as the dot product between two vectors, each representing the high dimensional vector mapping of graphs.
\begin{equation}
    \begin{split}
        k_{shortest\_path}(G_q,G_c)= \Phi(G_q) .\Phi(G_c)
    \end{split}
\end{equation}
\textbf{Normalised shortest path kernel}\cite{grakel}-
One can normalise the vectors $\Phi(G)$ and obtain the normalised shortest path kernel values.
\begin{equation}
    \begin{split}
        k_{Norm\_shortest\_path}(G_q,G_c)= \frac{\Phi(G_q) .\Phi(G_c)}{\left\lvert\Phi(G_q)\right\rvert_2 \left\lvert\Phi(G_c)\right\rvert_2}
    \end{split}
\end{equation}
\subsection{Reweighted Random Walk kernel \cite{rrwm}}

The main idea in a random walk kernel for graph similarity is to count the number of matching walks between the two graphs where the nodes are matched on the basis of their node labels. But one can obtain this kernel value in the by using the following formulation.
\begin{equation}
    \begin{split}
        k_{random\_walk}(G_q,C_c)=\sum\limits_{i,j=1}^{\lvert V_x \rvert} [\sum\limits_{n=0}^{N}\lambda_n A_x^n]_{ij}
    \end{split}
\end{equation}
But they are not effective in realf-life datasets when there are a lot of noisy and outlier nodes present in the association graph of two similar graphs. To solve this, a reweighted random walk kernel iteratively updates the confidence in the candidate correspondences between nodes of two graphs,  hopefully becoming robust against noisy and outlier nodes.

\section{Trainable Methods}

\subsection{GMN (Graph Matching Networks) \cite{gmn}}
This method outputs the similarity score by using a cross graph attention mechanism. This is in contrast to siamese networks where the two graphs are independently mapped to two vectors and then compared. In GMN, the information of how similar or dis-similar a node in one graph is with every other node in the other graph is also used while calculating the node embeddings.
\begin{figure}[!h]
    \centering
    \includegraphics[width=\textwidth]{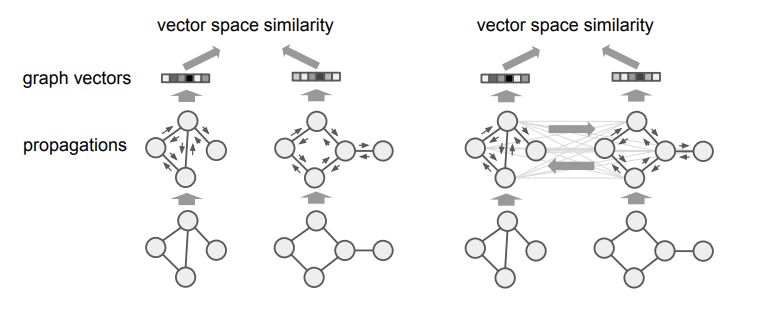}
    \caption{Left: Siamese network using graph embedding method, Right: GMN model}
    \label{fig:my_label}
\end{figure}

\chapter{Biased/ Trainable walks in GxNet}
\section{Introduction}
The SIGIR GxNet paper was based on Uniform Random walks to sample the subgraph within the product graph $\Gcal_{qc}$. In a scenario where the product graph is very huge there might be a situation where the uniform random walk completely misses the target subgraph resulting in a low score and thereby poor ranking. To avoid such a situation, we thought of training the walk paths instead of doing a random sampling. We changed the  start  probability  and  transition  probabilities  to  depend  on  node  or  edge features in  simple  parametric  ways. Consider  the  edge
 $(u,u′) \to (v,v′)$ in the product graph connecting $(u,u′)$ to $(v,v′)$. The corresponding node and edge features are $f_{(u,u')}$, $t_{(u,u'),(v,v')}$ and $f_{(v,v')}$. For random walk over the product graph, we designate the weight of the edge as $g_\phi(f_{(u,u')}, t_{(u,u'),(v,v')}, f_{(v,v')} )$. where $g_\phi$ is a neural network denoting a new set of non-linear parameters $\phi$. Edge weights are then suitably scaled per-node to make the walk probability matrix stochastic. 
 
 \section{Method}
 Following the notation of GxNet, in the trainable walks the weight of the edge is trained over a neural network whose parameters are denoted by $\phi$. Thus, in the trainable walks the pairwise ranking loss function would become:

\begin{align}
\Loss(G,C,C';\theta,\phi)= \sum_{G_q}\sum_{G_c,G_c'}\text{ReLU}(\Delta + y_{\theta,\phi}(G_q,G_{c'})-
y_{\theta,\phi}(G_q,G_c)) 
\end{align}

We break the down the training process in three possible ways:
\begin{itemize}
    \item Train both $\theta$ and $\phi$ simultaneously
    \item Train $\theta$ and $\phi$ alternatively
    \item Train $\phi$ first and then train $\theta$ while keeping the parameters of $\phi$ fixed
\end{itemize}

Along with the training procedure, we also experimented with the score aggregation function for a random walk across two formulations:
\begin{itemize}
    \item 
    MAX formulation: Selecting the max score across the $n$ walks of $k \epsilon [K]$ lengths 
    \begin{align}
        y_{\theta,\phi}(G_q, G_c, k)& = \max_{\substack{\Scal\sim P_{\phi}(\Scal|\Gcal^{qc}) |\Scal|= k }}(y_{\theta}(G_q,G_{c},S))\\
        &  \approx \frac{1}{\alpha} \log (  \sum_{ \substack{\Scal\sim P_{\phi}(\Scal|\Gcal^{qc}) |\Scal|= k }} \exp (\alpha y_{\theta}(G_q,G_{c},S)) ) 
        \\
        & =
        \frac{1}{\alpha} \log  ( N \times \sum_{S| |S| = k} \exp (\alpha \times y_{\theta}(G_q,G_{c},S)) \frac{P_{\phi}(\Scal|\Gcal^{qc})}{P_{\text{BiasedWalk}}(\Scal|\Gcal^{qc})} * P_{\text{BiasedWalk}}(\Scal|\Gcal^{qc}) )\\
        &
        = \frac{1}{\alpha} \log  ( \sum_{S\sim \text{BiasedWalk} |S| = k} \exp (\alpha \times y_{\theta}(G_q,G_{c},S)) \frac{P_{\phi}(\Scal|\Gcal^{qc})}{P_{\text{BiasedWalk}}(\Scal|\Gcal^{qc})} )
    \end{align}
    where $\alpha$ is a hyper-parameter
    \item
    AVERAGE formulation: Averaging the scores across $n$ walks of $k \epsilon [K]$ lengths.  \\
    For a fixed walk length k,
    \begin{align}
       y_{\theta,\phi}(G_q, G_c, k) & = \EE_{\substack{\Scal\sim P_{\phi}(\Scal|\Gcal^{qc}) \\ |\Scal|= k }}
      [y_{\theta}(G_q,G_{c},S)] \\ 
      & =\sum_{S : |S| = k} y_{\theta}(G_q,G_{c},S) P_{\phi}(\Scal|\Gcal^{qc}) \\
      & =\sum_{S : |S| = k} y_{\theta}(G_q,G_{c},S) \frac{P_{\phi}(\Scal|\Gcal^{qc})}{P_{\text{BiasedWalk}}(\Scal|\Gcal^{qc})}*P_{\text{BiasedWalk}}(\Scal|\Gcal^{qc}) \\
      & = \frac{1}{N} \times \sum_{\substack{S\sim \text{BiasedWalk} \\ |S| = k}} y_{\theta}(G_q,G_{c},S) \frac{P_{\phi}(\Scal|\Gcal^{qc})}{P_{\text{BiasedWalk}}(\Scal|\Gcal^{qc})}
    \end{align}
    Finally aggregating over walk lengths,
        \begin{align}
        y_{\theta, \phi}(G_q, G_c) = 
        \frac{1}{K} \sum\limits_{k \epsilon [K]} \lambda^k  y_{\theta, \phi} (G_q, G_c, k)  
        \end{align}
    where $\lambda$ is a hyper-parameter, we kept $\lambda$ = 1 unless specified.
\end{itemize}

\section{Results}
The trainable walks certainly beat the uniform walks in the performance. Though, when compared to other baselines, the trainable walks fall a little short. In the next section we tried to improve upon this gap by taking a new approach of shortest path product graphs.  

\begin{table}[]
\begin{tabular}{|l|l|l|l|l|l|l|}
\hline
\textbf{Dataset} &
  \textbf{\begin{tabular}[c]{@{}l@{}}GxNET\\ (Unbiased walks)\end{tabular}} &
  \textbf{\begin{tabular}[c]{@{}l@{}l@{}}GxNET\\ (Biased walks)\\ fixed $\phi$\end{tabular}} &
  \textbf{\begin{tabular}[c]{@{}l@{}l@{}}GxNET\\ (Biased walks) \\ trainable $\phi$\end{tabular}} &
  \textbf{GMN} &
  \textbf{RRW} &
  \textbf{SPK} \\ \hline
\textbf{\begin{tabular}[c]{@{}l@{}}VQA-Clevr\\ (0.10 noise)\end{tabular}} & 0.98 & \textbf{0.99} & \textbf{0.99} & 0.38 & 0.65          & 0.64 \\ \hline
\textbf{SQuAD}                                                        & 0.34 & 0.41          & \textbf{0.42} & 0.27 & 0.37          & 0.29 \\ \hline
\textbf{COX2}                                                         & 0.31 & 0.34          & 0.36           & 0.35 & \textbf{0.37} & 0.25 \\ \hline
\end{tabular}
\caption{Mean Average Precision values of various approaches. The best performing approaches are emphasised, suggesting superior performance of GxNET. GMN (Graph Matching Networks), RRW (Reweighted Random walk kernel) SPK (shortest path kernel) $\phi$ refers to the trainable parameters of learning transition probabilities in the random walks}
\label{tab:my-table}
\end{table}

\chapter{Neural matching with Shortest path product graph}

\section{Introduction}
After observing the performance of shortest path kernel on SQuAD dataset, we decided to include the idea of shortest path kernel in our product graph setting. The motive behind this was that in a random walk in a product graph of original graphs the maximum hop that the walk can reach is limited while in a product graph of shortest graphs, all nodes are within reach and the information of the distance is also retained. We call this approach Shortest Path Random Walk (SPRW). 

\section{Method}
Give a query graph $G_q$ and corpus graph $G_c$, we first convert it into its shortest path graphs $G_{qs}$ and $G_{cs}$ respectively. Note that the shortest graph is a complete graph with the edges being the distance of the two nodes in the original graphs. Thus evidently, if two nodes are not connected in the original graph then in the shortest graph they will be connected with a edge distance of $inf$ (infinity). 
Further, the tensor product of $G_{qs}$ and $G_{cs}$, $\Gcal_{qcs}$ was constructed. 

To test out the advantages and the potential of shortest product graphs we experimented with the biased walks by carefully curating the transition probability to first match with the shortest path kernel baseline and then surpass it in performance.

\paragraph{Start probability} Instead of a uniform random selection of the start node, we enforced the start nodes to be the matching nodes in the query and corpus graphs, i.e. 
  Give a start node $(u, u')$ in the product graph $\Gcal_{qcs}$, its start probability $P_{(u,u')}$ will be 
 \begin{align}
     P_{(u,u')} =  \frac{\mathbbm{1}(l(u)=l(u')) + \epsilon \mathbbm{1}(l(u) \neq l(u')) }{ \sum_{(u,u') \epsilon \Gcal_{qcs}}{ \mathbbm{1}(l(u)=l(u')) + \epsilon \mathbbm{1}(l(u) \neq l(u'))  }}
 \end{align}
 where $l(u)$ = $\mathtt{label}(u)$  and $l(u')$  = $\mathtt{label}(u')$ in the original graphs $G_{qs}$ and $G_{cs}$

\paragraph{Transition probability}  
 Similarly, transitions to the matching nodes with the same shortest distance in the query and corpus graphs was favoured over other transitions. Moreover, by tweaking the shortest distance limit we could vary our method from shortest path kernel to a more general setting. Formally, for a transition $(u,u') \to (v,v')$ the transition probability $P_{(u,u')\to(v,v')}$ is
 \begin{align}
     P_{(u,u')\to(v,v')} = \frac{\mathbbm{1}(l(u)=l(u')). \mathbbm{1}(l(v)=l(v')). \mathbbm{1}(\texttt{sp}(u,v)= \texttt{sp}(u', v')) .
     \mathbbm{1}(\texttt{sp}(u,v) \leq \alpha)}
     {\sum_{(u,u')\to (v,v') \epsilon \Gcal_{qcs}}\mathbbm{1}(l(u)=l(u')). \mathbbm{1}(l(v)=l(v')). \mathbbm{1}(\texttt{sp}(u,v)= \texttt{sp}(u', v')) .
     \mathbbm{1}(\texttt{sp}(u,v) \leq \alpha)} 
 \end{align}
 where $\texttt{sp}(u,v)$ is the shortest distance between nodes $u$ and $v$ in $G_{qs}$ and $\alpha$ is a hyper-parameter denoting the maximum distance of shortest path to consider for hopping. Hence, if $\alpha$ = $inf$, the above formulation reduces to the shortest path kernel. 
 
  Note that there is no training involved in SPRW.

\section{Results}

\begin{table}[!h]
\begin{tabular}{|l|l|l|l|l|l|l|}
\hline
\textbf{Dataset} &
  \textbf{\begin{tabular}[c]{@{}l@{}}Random \\ Prediction \end{tabular}} &
  \textbf{SPK} &
  \textbf{GMN} &
  \textbf{\begin{tabular}[c]{@{}l@{}}SPRW \\ (unbiased)\end{tabular}} &
  \textbf{\begin{tabular}[c]{@{}l@{}l@{}} SPRW\\ (biased) \\ $\texttt{SP}_{max}=1$\end{tabular}} &
  \textbf{\begin{tabular}[c]{@{}l@{}l@{}}SPRW\\ (biased) \\ $\texttt{SP}_{max}$=inf)\end{tabular}} \\ \hline
\textbf{MUTAG}   & 0.56 & 0.70 & \textbf{0.78} & 0.648 & 0.675 & 0.67 \\ \hline
\textbf{PTC\_MM} & 0.53 & 0.55 & \textbf{0.57} & 0.54 & 0.54 & 0.54 \\ \hline
\textbf{PTC\_FM} & 0.52 & 0.55 & \textbf{0.57} & 0.35 & ---  & ---  \\ \hline
\end{tabular}
\caption{Mean Average Precision values. The best performing approaches are emphasised, suggesting the superior performance of GMN. Note that GMN (Graph Matching Networks), RW (Random walk kernel) SPK (shortest path kernel), SPRW (Shortest Path Random Walk) has three settings (a) unbiased random walks (b) biased random walks with $\alpha$ = 1 and (c) biased random walks with $\alpha$ = $inf$ }
\label{tab:my-table-2}
\end{table}

\section{Conclusion}
From our research we draw following conclusions:
\begin{itemize}
    \item GxNet (biased walks with trainable $\phi$ ) performs better than GxNet (biased walks with fixed $\phi$ ) which in turn is better than GxNet based on uniform random walks (the original SIGIR paper) \cite{10.1145/3397271.3401216}
    \item In many cases normalised Shortest Path Kernel beats the Gxnet approach as was found with SQuAD dataset. Un-normalised Shortest Path Kernel still remains beaten by GxNet approach
    
    \item On TUDatasets, Shortest Path Kernel performs similar to Random Prediction denoting that it fails to understand the semantic relations in these dataset

\end{itemize}

\bibliographystyle{acm}
\bibliography{bibs/sample}

\end{document}